\documentclass[11pt]{article}

\usepackage{amsmath, amssymb}

\usepackage{amsthm}

\usepackage{ifthen}

\usepackage{ifpdf}

\ifpdf
\usepackage[pdftex]{graphicx}
\usepackage[pdftex]{hyperref}
\else
\usepackage[dvips]{graphicx}
\fi

\ifthenelse{\isundefined{\NoGenerateLetterSize}}
{
\ifpdf
\setlength{\pdfpagewidth}{8.5in}
\setlength{\pdfpageheight}{11in}
\else

\fi
}
{}

\newcommand{\Comment}[1]{}

\long\def\LongVersion#1\LongVersionEnd{#1}
\long\def\ShortVersion#1\ShortVersionEnd{}

\LongVersion 
\LongVersionEnd 

\ShortVersion 
\ShortVersionEnd 

\ShortVersion 
\usepackage{times}
\ShortVersionEnd 

\ShortVersion 
\renewcommand{\paragraph}[1]{\vspace{\parskip}\par\noindent\textbf{#1}}
\ShortVersionEnd 

\newtheorem{theorem}{Theorem}[section]
\newtheorem{lemma}[theorem]{Lemma}
\newtheorem{observation}[theorem]{Observation}
\newtheorem{corollary}[theorem]{Corollary}
\newtheorem{proposition}[theorem]{Proposition}

\theoremstyle{definition}
\newtheorem{property}[theorem]{Property}
\theoremstyle{plain}

\newtheorem{gtheorem}{Theorem}

\theoremstyle{definition}

\theoremstyle{plain}

\Comment{
\newtheorem{theorem}{Theorem}[section]
\newtheorem{lemma}{Lemma}[section]
\newtheorem{corollary}{Corollary}[section]

\newtheorem{proposition}{Proposition}[section]

\newtheorem{observation}{Observation}[section]

\theoremstyle{definition}

\theoremstyle{plain}
}

\newenvironment{AvoidOverfullParagraph}[0]
{\sloppy\ignorespaces}
{\par\fussy\ignorespacesafterend}

\newcommand{\Vertices}[0]{\mathit{V}}
\newcommand{\Edges}[0]{\mathit{E}}
\newcommand{\Weight}[0]{\mathit{w}}
\newcommand{\Reals}[0]{\mathbb{R}}
\newcommand{\Probability}[0]{\mathbb{P}}
\newcommand{\Expectation}[0]{\mathbb{E}}
\newcommand{\Variance}[0]{\mathrm{Var}}
\newcommand{\Xmax}[0]{x_{\max}}
\newcommand{\TailP}[0]{\pi}
\newcommand{\Approximation}[0]{\mathcal{A}}
\newcommand{\Instance}[0]{\mathcal{I}}
\newcommand{\Heavy}[0]{\mathit{H}}
\newcommand{\Light}[0]{\mathit{L}}
\newcommand{\ProcedureEstimate}[0]{\texttt{Estimate}}
\newcommand{\Polynomial}[0]{\mathrm{poly}}
\newcommand{\Estimator}[0]{\mathcal{E}}
\newcommand{\Failure}[0]{\mathrm{FAIL}}
\newcommand{\ProbZero}[0]{P_0}
\newcommand{\SharpP}[0]{{\#}P}
\newcommand{\AllHeavy}[0]{\mathcal{H}}
\newcommand{\Property}[0]{\mathcal{X}}
\newcommand{\Cost}[0]{\mathit{c}}

\newcommand{\RequirementScale}[0]{(\textrm{R1})}
\newcommand{\RequirementAdditive}[0]{(\textrm{R2})}
\newcommand{\RequirementContract}[0]{(\textrm{R3})}
\newcommand{\RequirementParallel}[0]{(\textrm{R4})}
\newcommand{\RequirementBounds}[0]{(\textrm{R5})}
\newcommand{\RequirementPositive}[0]{(\textrm{R6})}

\newcommand{\Distance}[0]{\mathrm{dist}}
\newcommand{\Diameter}[0]{\mathrm{diam}}
\newcommand{\Radius}[0]{\mathrm{rad}}
\newcommand{\MST}[0]{\mathrm{MST}}

\begin{document}

\title{Approximating the Statistics of various Properties in Randomly Weighted
Graphs}

\author{
Yuval Emek
\thanks{Microsoft Israel R\&D Center, Herzliya, Israel and
School of Electrical Engineering, Tel Aviv University, Tel Aviv, Israel.
E-mail: \texttt{yuvale@eng.tau.ac.il}.
Supported in part by The Israel Science Foundation, grant 664/05.}
\and
Amos Korman
\thanks{CNRS and Universit\'e Paris Diderot - Paris 7, France.
E-mail: \texttt{amos.korman@gmail.com}.
Supported in part by the ANR project ALADDIN, by the INRIA
project GANG, and by COST Action 295 DYNAMO.}
\and
Yuval Shavitt
\thanks{School of Electrical Engineering, Tel Aviv University, Tel Aviv,
Israel.
E-mail: \texttt{shavitt@eng.tau.ac.il}.}
}

\date{\today}

\maketitle

\begin{abstract}
Consider the setting of \emph{randomly weighted graphs}, namely, graphs whose
edge weights are chosen independently according to probability distributions
with finite support over the non-negative reals.
Under this setting, properties of weighted graphs typically become random
variables and we are interested in computing their statistical features.
Unfortunately, this turns out to be computationally hard for some properties
albeit the problem of computing them in the traditional setting of algorithmic
graph theory is tractable.
For example, there are well known efficient algorithms that compute the
\emph{diameter} of a given weighted graph, yet, computing the \emph{expected}
diameter of a given randomly weighted graph is \SharpP{}-hard even if the edge
weights are identically distributed.

In this paper, we define a family of properties of weighted graphs and show that
for each property in this family, the problem of computing the
\emph{$k^{\text{th}}$ moment} (and in particular, the expected value) of the
corresponding random variable in a given randomly weighted graph $G$ admits a
\emph{fully polynomial time randomized approximation scheme (FPRAS)} for every
fixed $k$.
This family includes fundamental properties of weighted graphs such as the
diameter of $G$, the \emph{radius} of $G$ (with respect to any designated
vertex) and the weight of a \emph{minimum spanning tree} of $G$.
\end{abstract}

\thispagestyle{empty}
\setcounter{page}{0}
\clearpage

\section{Introduction}
\label{section:Introduction}
In the traditional setting of graph algorithms, the input is typically a graph
$G$ whose edges are often associated with some \emph{weights}.
In most applications, $G$ represents a real-life network and the edge weights
correspond to some attributes of the network's links which are assumed to be
known when one wishes to apply some graph algorithm to $G$.
Unfortunately, in many scenarios these attributes of the real-life network
cannot be determined.
Still, however, it is often believed that the attributes of the network's
links are governed by some known probability distributions.
For example, the latency along each communication link in the Internet
backbone is usually assumed to be a random variable, rather than a fixed value
that can be determined a priori.
The same situation appears in the emerging field of networking called delay
tolerant networks (DTNs) \cite{Fall03} that includes sparse ah-hoc networks
\cite{ZKLTZ07,BCL07}, space exploration networks \cite{BHTFCDSW03}, submarine
networks \cite{PKL07}, and sensor networks.

Our goal in this paper is to advance the theoretic foundations of graph
algorithms operating in the context of edge weights that obey some specified
probability distributions.
In this context, various properties of the graph become random variables and
we wish to design better algorithms for the problems of computing the
statistical features of these random variables.
This turns out to be a non-trivial task even for basic and fundamental
graph properties which are easy to compute in the traditional (deterministic)
setting.
Similar challenges were previously tackled in many works (see the survey of
Ball et al. \cite{BCP95}), however the computational angle of these problems
received little treatment.

\paragraph{The model.}
\Comment{
\subsection{The model}
} 
A \emph{randomly weighted (RW) graph} is a graph\footnote{
Unless stated otherwise, all graphs in this paper are assumed to be finite and
undirected, though not necessarily simple.
The vertex set and edge set of graph $G$ are denoted by $\Vertices(G)$ and
$\Edges(G)$, respectively.
} $G$ in which the edge weights are independent random variables with finite
support over the non-negative reals.
Specifically, every edge $e \in \Edges(G)$ is associated with some positive
integer $m(e)$ and with some non-negative reals $W_{e}^{1}, \dots,
W_{e}^{m(e)}$ and $p_{e}^{1}, \dots, p_{e}^{m(e)}$, where $\sum_{i = 1}^{m(e)}
p_{e}^{i} = 1$, such that the weight of $e$ is $\Weight(e) = W_{e}^{i}$
with probability $p_{e}^{i}$ independently of all other edges.
The reals $W_{e}^{1}, \dots, W_{e}^{m(e)}$ are called the \emph{phases} of
edge $e$.

A special subclass of RW graphs that plays a major role in this paper is
that of \emph{biased coin weighted (BCW) graphs} in which $m(e) = 2$ for
all edges $e \in \Edges(G)$.
It will be convenient to slightly change the notation for BCW graphs:
edge $e \in \Edges(G)$ is said to take on its \emph{heavy} phase
$W_{e}^{\Heavy}$ with probability $p_e$ and on its \emph{light} phase
$W_{e}^{\Light}$ with probability $1 - p_e$, where $W_{e}^{\Heavy} \geq
W_{e}^{\Light} \geq 0$.
A BCW graph $G$ in which $p_e = p$, $W_{e}^{\Heavy} = W^{\Heavy}$,
and $W_{e}^{\Light} = W^{\Light}$ for all edges $e \in \Edges(G)$ is referred
to as an \emph{identically distributed weighted graph}.

\paragraph{Weighted graph properties.}
\Comment{
\subsection{Weighted graph properties}
} 
Let $\mathcal{G}$ be the collection of all (not necessarily simple)
connected\footnote{
Graph $G$ is said to be connected if it admits a path between $u$ and $v$ for
every two vertices $u, v \in \Vertices(G)$.
} graphs $G$ with non-negative edge weights $\Weight : \Edges(G) \rightarrow
\Reals_{\geq 0}$.
Throughout, we think of a \emph{weighted graph property} as a function
$\Property$ that assigns a non-negative real $\Property(G)$ to each graph $G
\in \mathcal{G}$.
A weighted graph property $\Property : \mathcal{G} \rightarrow \Reals_{\geq
0}$ is said to be \emph{distance-cumulative} if it satisfies the following
requirements for every graph $G \in \mathcal{G}$: \\
\textbf{\RequirementScale{}}
If $G' \in \mathcal{G}$ is the graph obtained from $G$ by multiplying all edge
weights by some factor $r \in \Reals_{\geq 0}$, then $\Property(G') = r \cdot
\Property(G)$. \\
\textbf{\RequirementAdditive{}}
If $G' \in \mathcal{G}$ is the graph obtained from $G$ by increasing the
weight of some edge $e \in \Edges(G)$ by an additive term $r \in \Reals_{\geq
0}$, then $\Property(G) \leq \Property(G') \leq \Property(G) + r$. \\
\textbf{\RequirementContract{}}
If $\Weight(e) = 0$ for some edge $e \in \Edges(G)$, then $\Property(G / e) =
\Property(G)$, where $G / e \in \mathcal{G}$ is the graph obtained from $G$ by
contracting the edge $e$. \\
\textbf{\RequirementParallel{}}
If $e, e' \in \Edges(G)$ are parallel edges and $\Weight(e) \geq \Weight(e')$,
then $\Property(G - e) = \Property(G)$, where $G - e \in \mathcal{G}$ is the
graph obtained from $G$ by removing the edge $e$. \\
\textbf{\RequirementBounds{}}
If $\Property(G)$ is strictly positive, then $\Property(G) \geq \min\{
\Weight(e) \mid e \in \Edges(G) \}$. \\
\textbf{\RequirementPositive{}}
$\Property(G) = 0$ if and only if the graph obtained from $G$ by removing all
(strictly) positive weight edges is connected.

Requirements~\RequirementScale{}--\RequirementBounds{} and the sufficiency
direction of requirement~\RequirementPositive{} are naturally satisfied by
weighted graph properties in which the edge weights correspond to time delays,
routing costs, etc.
The necessity direction of requirement~\RequirementPositive{} is slightly more
restrictive.
We extend the definition of distance-cumulative weighted graph properties
$\Property$ by assuming that $\Property(G) = \infty$ for every disconnected
graph $G$.
A weighted graph property $\Property$ is said to be \emph{efficiently
calculated} if there exists a polynomial time algorithm that computes
$\Property(G)$ for every graph $G$.

\ShortVersion 
\begin{AvoidOverfullParagraph}
\ShortVersionEnd 
Consider some graph $G \in \mathcal{G}$.
The \emph{diameter} of $G$, denoted $\Diameter(G)$, is defined as
$\Diameter(G) = \max\{ \Distance_{G}(u, v) \mid u, v \in \Vertices(G) \}$,
where $\Distance_{G}(u, v)$ is the \emph{distance} between $u$ and $v$ in $G$,
namely, the length of a shortest (with respect to the edge weights) path from
$u$ to $v$.
For a designated vertex $u \in \Vertices(G)$, the \emph{radius} of $G$,
denoted $\Radius(G)$, is defined as $\Radius(G) = \max\{ \Distance_{G}(u, v)
\mid v \in \Vertices(G) \}$.
The weight of a \emph{minimum spanning tree} of $G$, denoted $\MST(G)$, is
defined as $\MST(G) = \min\{ \sum_{e \in T} \Weight(e) \mid T \text{ is a
spanning tree of } G \}$.
It is easy to verify that $\Diameter(G)$, $\Radius(G)$, and $\MST(G)$ are
distance-cumulative weighted graph properties.
Efficient algorithms that compute them are described in most textbooks on
graph algorithms (e.g., \cite{Even79}).
Another distance-cumulative weighted graph property is the diameter of a
\emph{min-diameter spanning tree}, defined as $\min\{ \Diameter(T) \mid T
\text{ is a spanning tree of } G \}$.
The min-diameter spanning tree is also known to be efficiently calculated
\cite{HT95}.
\ShortVersion 
\end{AvoidOverfullParagraph}
\ShortVersionEnd 

\ShortVersion 
\begin{AvoidOverfullParagraph}
\ShortVersionEnd 
Two more efficiently calculated weighted graph properties that fit into the
definition of distance-cumulative properties by slightly modifying
requirement~\RequirementBounds{} are the \emph{all pairs average distance} and
the \emph{single source average distance} with respect to some designated
vertex.
Our techniques can be adjusted to handle such a modification, however, this is
beyond the scope of the current version of the paper.
\ShortVersion 
\end{AvoidOverfullParagraph}
\ShortVersionEnd 

\paragraph{Our contribution.}
\Comment{
\subsection{Our contribution}
} 
Let $\Property$ be some distance-cumulative weighted graph property and assume
that it is efficiently calculated.
Given some connected RW graph $G$, $\Property(G)$ is a random variable ---
denote it by $X$ --- and we are interested in approximating its
\emph{$k^{\text{th}}$ moment}, i.e., $\Expectation[X^{k}]$, for any fixed
$k$.
Specifically, we develop a \emph{fully polynomial time randomized
approximation scheme (FPRAS)} for the problem, namely, a randomized algorithm
that runs in time $\Polynomial(|G|, 1 / \epsilon)$ for any choice of $\epsilon
> 0$, where $|G|$ stands for the number of bits required to encode $G$ in a
standard binary representation, and returns a $(1 + \epsilon)$-approximation
of $\Expectation[X^{k}]$ with probability\footnote{
Using the \emph{median of means method}, the success probability of an FPRAS
can be increased to $1 - \hat{\epsilon}$ for any choice of $\hat{\epsilon} >
0$ at the cost of increasing the run-time by an $O (\log (1 /
\hat{\epsilon}))$ factor.
} at least $3 / 4$.
The following theorem is established in
\LongVersion 
Sections \ref{section:Fpras}, \ref{section:ProcedureEstimate}, and
\ref{section:TransformingRwToBcw}.
\LongVersionEnd 
\ShortVersion 
Sections \ref{section:Fpras} and \ref{section:ProcedureEstimate} and in
Appendix~\ref{section:TransformingRwToBcw}.
\ShortVersionEnd 

\begin{gtheorem} \label{gtheorem:Fpras}
The problem of computing the $k^{\text{th}}$ moment of $\Property(G)$ on
connected RW graphs $G$ admits an FPRAS for every fixed $k$.
\end{gtheorem}

In general, Theorem~\ref{gtheorem:Fpras} is best possible.
This is because exact solutions to the problems of computing
$\Expectation[\Diameter(G)]$ and $\Expectation[\Radius(G)]$ are \SharpP{}-hard
to obtain even when the input is restricted to identically distributed
weighted graphs.
Refer to
\LongVersion 
Section~\ref{section:Hardness}
\LongVersionEnd 
\ShortVersion 
Appendix~\ref{section:Hardness}
\ShortVersionEnd 
for a proof of this rather simple observation.

To the best of our knowledge, our FPRAS yields the first provably polynomial
time algorithm with guaranteed approximation ratio for any non-trivial
statistical feature of a weighted graph property in randomly weighted graphs.
Moreover, it seems that most related literature focuses on individual weighted
graph properties and does not attempt to provide a framework for a more general
theory of such properties;
indeed, Snyder \& Steele call for such a framework in their survey
\cite{SS95}.
We hope that our technique which is suitable for all distance-cumulative
weighted graph properties will be a significant step in that direction.

\paragraph{Related work.}
\Comment{
\subsection{Related work}
} 
The algorithmic aspects of randomly weighted graphs have been extensively
studied since the early 60's (cf. Fulkerson \cite{Fulk62}) mainly in the
context of the shortest $(s, t)$-path, the longest $(s, t)$-path (a.k.a. the
PERT problem), and maximum $(s, t)$-flow.
A comprehensive account of the various methods developed for the computation
(and approximation) of the statistical features corresponding to these
weighted graph properties is provided by Ball et al. \cite{BCP95} who also
observe that an exact computation of the expected values of these weighted
graph properties is \SharpP{}-hard.
Note that except for a few special cases (e.g., series-parallel networks with
a specific type of edge distributions), none of the algorithms developed in
that context is provably polynomial with guaranteed approximation ratio.

Distance-cumulative weighted graph properties were also investigated under
this setting.
Hassin \& Zemel \cite{HZ85} prove that the diameter (and radius) of a complete
$n$-vertex graph whose edge weights are uniformly and independently
distributed in $[0,1]$ is almost surely $\Theta (\log(n) / n)$.
The hidden constants in this expression were resolved by Janson
\cite{Jans99} who shows that the $(s, t)$-distance, radius with respect to
$s$, and diameter of a complete $n$-vertex graph whose edge weights are
uniformly and independently distributed in $[0,1]$ converge in probability to
$\ln n / n$, $2 \ln n / n$, and $3 \ln n / n$, respectively.
Frieze \cite{Frie85} shows that for every distribution function $F$ with finite
variance whose derivative at zero exists and satisfies $F'(0) = D > 0$, if the
edge weights in a complete $n$-vertex graph $G$ are independently distributed
according to $F$, then the weight of a minimum spanning tree of $G$ converges
in probability to $\zeta(3) / D$, where $\zeta(3) = \sum_{j = 1}^{\infty} 1 /
j^3$.
This is generalized by Steele \cite{Stee87} who shows that the assumption on
the finite variance can be lifted.
For the special case of $F$ being the uniform distribution over $[0, 1]$,
Beveridge et al. \cite{BFM98} establish a variant of this bound for
$r$-regular graphs.

Asymptotic results for minimum spanning trees on $n$ points uniformly and
independently distributed in the Euclidean unit ball are established by
Bertsimas \& van Rysin \cite{BR90}.
Kulkarni \cite{Kulk88} and Alexopoulos \& Jacobson \cite{AJ00} present
algorithms that compute the distribution of $\MST(G)$ for graphs $G$ whose
edge weights obey exponential and discrete distributions, respectively.
The run-times of their algorithms are not necessarily polynomial, though.
A non-trivial upper bound on $\Expectation[\MST(G)]$ is established by Jain \&
Mamer \cite{JM88}.

A different, yet, related subject that admits a plethora of literature is
average case analysis for graph algorithms
(e.g. \cite{FG85,KKP93,CFMP00,Maye03}).
There it is assumed that the edge weights in the input of some graph algorithm
are drawn from a specified probability distribution and the goal is to analyze
the expected run-time of the algorithm with respect to that distribution;
refer to \cite{FM97} for a survey.
It is important to point out that upon invocation of the graph algorithm, the
actual edge weights are determined, and in particular, known to the algorithm,
in contrast to the RW graphs setting in which the challenge is to cope with
the uncertainty in the edge weights.

\paragraph{Techniques.}
\Comment{
\subsection{Techniques}
} 
We now provide an informal overview of the construction of an FPRAS for
$\Expectation[\Property(G)]$, where $\Property$ is some efficiently calculated
distance-cumulative weighted graph property and $G$ is an RW graph;
approximating higher moments is very similar.
Note that if the variance of $\Property(G)$ is at most some polynomial times
$\Expectation[\Property(G)]^2$ (that is, the critical ratio is polynomial),
then $\Expectation[\Property(G)]$ can be approximated by means of sampling
(a.k.a. Monte Carlo method).
However, there exist some simple examples
(cf. Appendix~\ref{appendix:CriticalRatio}) in which the variance is too
large, and therefore a different approach must be sought.

Our first step is to employ a simple reduction that allows us to focus on BCW
graphs rather than arbitrary RW graphs.
This reduction is presented in
\LongVersion 
Section~\ref{section:TransformingRwToBcw}.
\LongVersionEnd 
\ShortVersion 
Appendix~\ref{section:TransformingRwToBcw}.
\ShortVersionEnd 
So, in what follows our goal is to approximate $\Expectation[\Property(G)]$ to
within a multiplicative error of $1 + O (\epsilon)$, where $G$ is a BCW graph.

The desired approximation would have been straightforward to obtain if we
could have efficiently approximated $\Probability(\Property(G) > x)$ for
arbitrary choices of real $x \geq 0$.
It turns out that the case $x = 0$ is well known:
approximating $\Probability(\Property(G) > 0)$ is equivalent to approximating
the all-terminal network reliability problem (cf. \cite{BCP95}) in the graph
$G^0$ obtained from $G$ by removing all edges whose light phases are strictly
positive.
The FPRAS developed by Karger in \cite{Karg99} for the all-terminal network
reliability problem is used to obtain this approximation.
Karger's technique is based on identifying a collection $\mathcal{C}$ of
polynomially many ($2$-way) cuts in $G^0$ such that the probability that all
edges of some cut not in $\mathcal{C}$ take on their heavy phases is small.
(Of course, Karger uses the language of network reliability, rather than that
of BCW graphs.)

Unfortunately, approximating $\Probability(\Property(G) > x)$ to within a
multiplicative error of $1 + O (\epsilon)$ for an arbitrary choice of real $x
> 0$ seems to be a challenging task.
In fact, when $\Property = \MST$ (namely, we are required to approximate the
expected weight of a minimum spanning tree), an efficient implementation of
this task would yield an FPRAS for the Tutte polynomial $T_{G}(x, y)$ of
arbitrary graphs $G$ for every $x, y > 1$ (refer to Bollob\'{a}s \cite{Boll98}
for a comprehensive treatment of the Tutte polynomial and its many
applications);
whether or not such an FPRAS exists is an important open question
\cite{AFW95,GJ08}.

Instead, we use a careful ``sliding window'' argument to show that the desired
approximation of $\Expectation[\Property(G)]$ can be efficiently obtained by
repeatedly calling a procedure called Procedure~\ProcedureEstimate{}.
Given a positive real $x$, Procedure~\ProcedureEstimate{} approximates
$\Probability(\Property(G) > x)$ to within an additive error of $O (\epsilon)
\cdot \Probability(\Property(G) > 0)$.
In other words, we show that it suffices to produce a weaker approximation of
$\Probability(\Property(G) > x)$;
the quality of this weaker approximation is determined by
$\Probability(\Property(G) > 0)$.
The ``sliding window'' argument, presented in Section~\ref{section:Fpras}, is
based on iteratively resetting all edge phases in $G$ smaller than some
threshold for a carefully chosen sequence of thresholds.

Procedure~\ProcedureEstimate{} itself is presented in
Section~\ref{section:ProcedureEstimate}.
The main trick there relies on formulating the real valued random variable $Y$
that maps each instance $\Instance$ of the probability space defined by
$\Edges(G) - \Edges(G^0)$ to $\Probability(\Property(G) > x \mid \Instance)$.
Since $\Expectation[Y] = \Probability(\Property(G) > x)$, it is sufficient
to approximate $\Expectation[Y]$.
We would have wanted to do so by sampling instances $\Instance$ of the
probability space defined by $\Edges(G) - \Edges(G^0)$ and then computing
$\Probability(\Property(G) > x \mid \Instance)$.
Sampling instances of the probability space defined by $\Edges(G) -
\Edges(G^0)$ is a straightforward task.
The problem is that given such an instance $\Instance$, it is not clear how to
efficiently compute $\Probability(\Property(G) > x \mid \Instance)$.

To tackle this obstacle, we revisit the cut collection $\mathcal{C}$ and for
each instance $\Instance$, identify those cuts in $\mathcal{C}$ that
conditioned on $\Instance$, imply $\Property(G) > x$.
For that to work, we must extend Karger's construction of $\mathcal{C}$ to
$r$-way cuts for all $r \geq 2$.
This extension builds upon the recent bound of Berend \& Tassa on the Bell
number \cite{BT10}.
We then employ the method of Karp, Luby, and Madras \cite{KL85,KLM89} for
probabilistic DNF satisfiability to approximate the probability
that at least one of these cuts is induced by the edges in $\Edges(G^0)$ that
take on their heavy phase.

\section{Preliminaries}
\label{section:Preliminaries}

\LongVersion
\begin{AvoidOverfullParagraph}
\LongVersionEnd
\paragraph{Randomly weighted graphs.}
\Comment{
\subsection{Randomly weighted graphs}
} 
Throughout we consider some distance-cumulative weighted graph property
$\Property{}$ and some $n$-vertex connected RW graph $G$.
Let $X$ denote the random variable that takes on $\Property(G)$.
By requirement~\RequirementScale{}, we may assume without loss of generality
that the edge phases of $G$ are scaled so that the smallest non-zero phase is
exactly $1$.
Consequently requirement~\RequirementBounds{} implies that $X$ is either $0$,
or it is bounded from below by $1$.
On the other hand, requirements \RequirementAdditive{} and
\RequirementParallel{} guarantee that $X$ is bounded from above by $\Xmax =
n^2 \cdot \max\{ W_{e}^{i} \mid e \in \Edges(G) \text{ and } 1 \leq i \leq
m(e) \}$.
\LongVersion
\end{AvoidOverfullParagraph}
\LongVersionEnd

Suppose that $G$ is a BCW graph.
In what follows we assume that $0 < p_e < 1$ for every edge $e \in \Edges(G)$
(this assumption is clearly without loss of generality as the phases of an
edge are not required to be disjoint).
Let $F \subseteq \Edges(G)$ be some subset of the edges.
Each edge $e \in F$ takes on its heavy phase with probability $p_e$ and on its
light phase with probability $1 - p_e$; this defines a probability space.
It will be convenient to view an \emph{instance} $\Instance$ of this
probability space as a Boolean function $\Instance : F \rightarrow \{\Heavy,
\Light\}$, where
\LongVersion 
$$
\Instance(e) =
\left\{
\begin{array}{ll}
\Heavy & \text{if $e$ takes on its heavy phase $W_{e}^{\Heavy}$;} \\
\Light & \text{if $e$ takes on its light phase $W_{e}^{\Light}$.}
\end{array}
\right.
$$
\LongVersionEnd 
\ShortVersion 
$\Instance(e) = \Heavy$ if $e$ takes on its heavy phase $W_{e}^{\Heavy}$; and
$\Instance(e) = \Light$ if $e$ takes on its light phase $W_{e}^{\Light}$.
\ShortVersionEnd 
At the risk of abusing notation, we may sometime write $F$ when we actually
refer to the probability space it defines; our intentions will be clear from
the context.

\paragraph{Cuts and compact cuts.}
\Comment{
\subsection{Cuts and compact cuts}
} 
Consider some connected graph $G$.
An \emph{$r$-way cut} $C$ of $G$ is a partition of $\Vertices(G)$ to
$r$ pairwise disjoint subsets, that is, $C = \{U_1, \dots, U_r\}$, where $
\bigcup_{1 \leq i \leq r} U_i = \Vertices(G)$ and $U_i \cap U_j =
\emptyset$ for every $i \neq j$.
The subsets $U_1, \dots, U_r$ are referred to as the \emph{clusters} of
$C$.
A \emph{cut} refers\footnote{
In some literature, a cut refers to a $2$-way cut, while an $r$-way cut for $r
> 2$ is called a \emph{multiway} cut.
We do not make this distinction.
} to an $r$-way cut for any $r \geq 2$.

Consider some $r$-way cut $C = \{U_1, \dots, U_r\}$ of $G$.
We say that an edge $e \in \Edges(G)$ \emph{crosses} $C$ if $e \in U_i
\times U_j$ for some $i \neq j$.
The set of edges crossing $C$ is denoted by $\Edges(C)$.
The cardinality $|\Edges(C)|$ is referred to as the \emph{size} of $C$;
if the edges of $G$ are assigned with positive \emph{costs} $\Cost :
\Edges(G) \rightarrow \Reals_{> 0}$, then the sum $\sum_{e \in \Edges(C)}
\Cost(e)$ is referred to as the \emph{cost} of $C$.
The cut $C$ is called \emph{compact} if $G(U_i)$ is connected for every $
1 \leq i \leq r$.
Note that every $r$-way cut is a compact $r'$-way cut for some $r' \geq r$.
A \emph{min cut} (respectively, \emph{min cost cut}) is a cut of minimum
size (resp., cost).
It is easy to verify that a min cut (resp., min cost cut) must be a compact
$2$-way cut.

Consider some subset $F \subseteq \Edges(G)$ and a compact cut $C$ of $G$.
We say that $F$ \emph{induces} the cut $C$ if the connected components of the
graph obtained from $G$ by removing the edges in $F$ agree with the clusters
of $C$.
In particular, $F$ must be a superset of $\Edges(C)$;
it may contain additional edges as long as the removal of these edges does not
disconnect any cluster of $C$.

\paragraph{Analogy to the all-terminal network reliability problem.}
\Comment{
\subsection{Analogy to the all-terminal network reliability problem}
} 
Consider some BCW graph $G$ and some distance-cumulative weighted graph
property $\Property$.
Suppose that we wish to approximate the probability that $\Property(G) > 0$.
Let $E^0 = \{ e \in \Edges(G) \mid W_{e}^{\Light} = 0 \}$ and let $G^0$ be the
restriction of $G$ to the edges in $E^0$.
Requirement~\RequirementPositive{} implies that the event $\Property(G) > 0$
depends only on the probability space $E^0$;
specifically, $\Property(G) > 0$ if and only if $\Property(G^0) > 0$.
Moreover, if $G^0$ is disconnected, then $\Property(G^0) = \infty$ with
probability $1$, thus we subsequently assume that $G^0$ is connected.
By employing requirement~\RequirementPositive{} once more, we conclude that
$\Property(G^0) > 0$ if and only if the edges that take on their heavy
(positive) phases under the probability space $E^0$ induce a cut on $G^0$.
This leads us to an analogy between the problem of approximating
$\Probability(\Property(G) > 0)$ and the \emph{all-terminal network
reliability (ATNR)} problem.

The input to ATNR is a connected undirected graph $H$ in which each edge $e$
\emph{fails} (i.e., removed) with some specified probability.
The goal is to compute the probability, referred to as the \emph{failure}
probability of $H$ and denoted $\Failure(H)$, that $H$ becomes disconnected
following such an edge failure experiment.
ATNR is known to be $\sharp$P-complete \cite{Vali79,BCP95} and Karger develops
an FPRAS for it \cite{Karg99}.
Since $H$ becomes disconnected if and only if the failing edges induce a cut
on it, we conclude that $\Probability(\Property(G^0) > 0) = \Failure(G^0)$,
where the latter is defined over an instance of ATNR in which each edge $e \in
\Edges(G^0)$ fails with probability $p_e$ (the probability that $e$ takes on
its heavy phase in the BCW graph framework).
Consequently, approximating $\Probability(\Property(G) > 0)$ can be performed
by a direct application of Karger's algorithm.

\paragraph{Monte Carlo method and approximators.}
\Comment{
\subsection{Monte Carlo method and approximators}
} 
Consider some probability space with sample space $\Omega$ and let $X : \Omega
\rightarrow \Reals$ be a real valued random variable over this probability
space.
Suppose that the expectation of $X$ is defined and denote it by $\mu$.
Let $X_1, \dots, X_n$ be $n$ independent samples of $X$ and fix $\bar{X} =
\sum_{i = 1}^{n} X_i / n$.
Evaluating $\mu$ by $\bar{X}$ is referred to as the \emph{Monte Carlo method}
(cf. \cite{KL85}).
Let $\epsilon$ and $\hat{\epsilon}$ be some positive reals.
The following two theorems are direct consequences of Chernoff's inequality
\cite{Cher52} (Theorem~\ref{theorem:MonteCarloChernoff}) and Hoeffding's
inequality \cite{Hoef63} (Theorem~\ref{theorem:MonteCarloHoeffding}).

\begin{theorem} \label{theorem:MonteCarloChernoff}
If $X$ is an indicator random variable (namely, $X \in \{0, 1\}$), then
taking $n \geq 4 \ln(2 / \hat{\epsilon}) \mu / \epsilon^2$ samples
guarantees that $\Probability(|\bar{X} - \mu| > \epsilon) \leq
\hat{\epsilon}$.
\end{theorem}

\begin{theorem} \label{theorem:MonteCarloHoeffding}
If $X$ is almost surely in the interval $[a, b]$, where $b - a = \rho$, then
taking $n \geq \ln(2 / \hat{\epsilon}) \rho^2 / (2 \epsilon^2)$ samples
guarantees that $\Probability(|\bar{X} - \mu| > \epsilon) \leq
\hat{\epsilon}$.
\end{theorem}

This leads us to notion of \emph{approximators}.
Consider some non-negative real value $v$ that we would like to approximate
with the real $v'$.
Then $v'$ is said to be an \emph{$(\epsilon, \hat{\epsilon})$-approximator} of
$v$ if it satisfies the inequality $|v - v'| \leq \epsilon$ with probability
at least $1 - \hat{\epsilon}$, where the probability is taken over the
randomness used to generate $v'$.
Under this terminology, Theorems \ref{theorem:MonteCarloChernoff} and
\ref{theorem:MonteCarloHoeffding} provide sufficient conditions to guarantee
that $\bar{X}$ is an $(\epsilon, \hat{\epsilon})$-approximator of $\mu$.

\begin{proposition} \label{proposition:ApproximatorsTransitivity}
If $v_1$ is an $(\epsilon_1, \hat{\epsilon_1})$-approximator of $v_0$ and
$v_2$ is an $(\epsilon_2, \hat{\epsilon_2})$-approximator of $v_1$, then
$v_2$ is an $(\epsilon_1 + \epsilon_2, \hat{\epsilon_1} +
\hat{\epsilon_2})$-approximator of $v_0$.
\end{proposition}

\begin{proposition} \label{proposition:ApproximatorsAdditivity}
If $v'_i$ is an $(\epsilon, \hat{\epsilon})$-approximator of $v_i$ for every $1
\leq i \leq n$, then $\sum_{i = 1}^{n} v'_i / n$ is an $(\epsilon, n \cdot
\hat{\epsilon})$-approximator of $\sum_{i = 1}^{n} v_i / n$.
\end{proposition}

\section{An FPRAS for BCW graphs}
\label{section:Fpras}
In this section we consider some $n$-vertex BCW graph $G$ and some small
performance parameter $\epsilon > 0$;
our goal is to approximate $\Expectation[X^{k}]$ to within a multiplicative
error of $1 + O (\epsilon)$.
Here, we restrict our attention to the case $k = 1$, that is, we approximate
$\Expectation[X]$.
Extending our result to larger (yet fixed) values of $k$ is mainly a matter of
notation and we omit it from this version of the paper.
The approximation presented in this section builds upon the more sophisticated
Procedure~\ProcedureEstimate{} which is presented in
Section~\ref{section:ProcedureEstimate}.

\begin{theorem} \label{theorem:FprasBcw}
There exists a randomized algorithm that with probability at least $3 / 4$,
approximates $\Expectation[X]$ to within a multiplicative error of $1 + O
(\epsilon)$ in time $\Polynomial(|G|, 1 / \epsilon)$.
\end{theorem}

Let $N$ be the smallest integer such that $\Xmax < (1 + \epsilon)^{N}$.
(Note that $N$ is proportional to $\log (\Xmax) / \epsilon = \Polynomial (|G|,
1 / \epsilon)$).
Clearly, we have $0 \leq X < (1 + \epsilon)^{N}$.
Fix $\TailP_i = \Probability(X \geq (1 + \epsilon)^{i})$ for every $0 \leq i
\leq N$.
Towards the approximation of $\Expectation[X]$, we first define
\LongVersion
\begin{align}
\Approximation
= & ~ \sum_{i = 1}^{N} (1 + \epsilon)^{i - 1} \cdot \Probability \left((1 +
\epsilon)^{i - 1} \leq X < (1 + \epsilon)^{i} \right) \nonumber \\
= & ~ \sum_{i = 1}^{N} (1 + \epsilon)^{i - 1} \cdot \left[ \TailP_{i - 1} -
\TailP_{i} \right] \nonumber \\
= & ~ \TailP_0 + \sum_{i = 1}^{N - 1} \epsilon \cdot (1 + \epsilon)^{i - 1}
\cdot \TailP_i - (1 + \epsilon)^{N - 1} \cdot \TailP_{N} \nonumber \\
= & ~ \TailP_0 + \sum_{i = 1}^{N - 1} \epsilon \cdot (1 + \epsilon)^{i - 1}
\cdot \TailP_i ~ , \label{equation:ExpressionApproximation}
\end{align}
\LongVersionEnd
\ShortVersion
\begin{align}
\Approximation
= & ~ \sum_{i = 1}^{N} (1 + \epsilon)^{i - 1} \cdot \Probability \left((1 +
\epsilon)^{i - 1} \leq X < (1 + \epsilon)^{i} \right)
= ~ \sum_{i = 1}^{N} (1 + \epsilon)^{i - 1} \cdot \left[ \TailP_{i - 1} -
\TailP_{i} \right] \nonumber \\
= & ~ \TailP_0 + \sum_{i = 1}^{N - 1} \epsilon \cdot (1 + \epsilon)^{i - 1}
\cdot \TailP_i - (1 + \epsilon)^{N - 1} \cdot \TailP_{N}
= ~ \TailP_0 + \sum_{i = 1}^{N - 1} \epsilon \cdot (1 + \epsilon)^{i - 1}
\cdot \TailP_i ~ , \label{equation:ExpressionApproximation}
\end{align}
\ShortVersionEnd
where (\ref{equation:ExpressionApproximation}) is due to the fact that $
\TailP_{N} = 0$.
It is easy to verify that
\begin{equation} \label{equation:ApproximationBound1}
\Expectation[X] / (1 + \epsilon)
\leq \Approximation
\leq \Expectation[X] ~ ,
\end{equation}
so our next goal is to approximate $\Approximation$.
Note that (\ref{equation:ExpressionApproximation}) enables the computation of
a $(1 + \epsilon)$-approximation of $\Expectation[X]$ based on $(1 +
\epsilon)$-approximations of $\Probability(X \geq x)$ for sufficiently many
values of $x$.
Unfortunately, we do not know how to obtain such an approximation directly and
we are forced to apply some modifications to $G$.

\paragraph{The shrunk graphs.}
\Comment{
\subsection{The shrunk graphs}
} 
Fix $\kappa = \left\lceil \log_{1 + \epsilon} \left( \frac{n^2 (1 +
\epsilon)}{\epsilon} \right) \right\rceil$.
For $i = 0, 1, \dots, N - 1$, let $G_i$ be the BCW graph obtained from $G$ by
setting $W_{e}^{\varphi} \leftarrow 0$ for every edge $e \in \Edges(G)$ and
phase $\varphi \in \{\Heavy, \Light\}$ such that $W_{e}^{\varphi} < (1 +
\epsilon)^{i - \kappa}$.
We refer to a phase that was set to $0$ in this process as a \emph{shrunk}
phase;
the graphs $G_0, \dots, G_{N - 1}$ are called the \emph{shrunk} graphs.
(Clearly, if the heavy phase of some edge is shrunk, then so is the
light phase.)

Let $X_i$ be the random variable that takes on $\Property(G_i)$.
Requirement~\RequirementAdditive{} guarantees that $X_0 \geq X_1 \geq \cdots
\geq X_{N - 1}$.
The assumption that the minimum positive phase is scaled to $1$ implies that
$G = G_0 = G_1 = \cdots = G_{\kappa}$, hence $X = X_0 = X_1 = \cdots =
X_{\kappa}$.
\LongVersion 
If $i > \kappa$, then $X_i$ may be smaller than $X$, however it is not much
smaller as depicted in the following proposition.
\LongVersionEnd 
\ShortVersion 
If $i > \kappa$, then $X_i$ may be smaller than $X$, however it is not much
smaller as depicted in the following proposition (proof deferred to
Appendix~\ref{appendix:AdditionalProofsFpras}).
\ShortVersionEnd 

\begin{proposition} \label{proposition:ShrinkingShortEdges}
If $X \geq (1 + \epsilon)^{i}$ for some $0 \leq i \leq N - 1$, then $X / (1 +
\epsilon) < X_i \leq X$.
\end{proposition}
\newcommand{\ProofPropositionShrinkingShortEdges}{
Since phase $W_{e}^{\varphi}$ shrinks in $G_i$ only if $W_{e}^{\varphi} < (1 +
\epsilon)^{i - \kappa} \leq (1 + \epsilon)^{i} \cdot
\frac{\epsilon}{n^2 (1 + \epsilon)}$,
requirements \RequirementAdditive{} and \RequirementParallel{} guarantee that
$$
X_i
> X - n^2 \cdot \frac{\epsilon (1 + \epsilon)^{i}}{n^2 (1 + \epsilon)}
> X - \frac{\epsilon X}{1 + \epsilon}
= X / (1 + \epsilon) ~ .
$$
The assertion follows.
} 
\LongVersion 
\begin{proof}
\ProofPropositionShrinkingShortEdges{}
\end{proof}
\LongVersionEnd 

\newcommand{\DetailsEquationApproximationBoundTwo}{
\begin{align*}
\frac{\Approximation}{(1 + \epsilon)}
~ = ~ & \frac{\TailP_0}{(1 + \epsilon)} + \sum_{i = 1}^{N - 1} \epsilon \cdot
(1 + \epsilon)^{i - 2} \cdot \TailP_i \\
\leq ~ & \frac{\TailP'_0}{(1 + \epsilon)} + \sum_{i = 1}^{N - 1} \epsilon \cdot
(1 + \epsilon)^{i - 2} \cdot \TailP'_{i-1} \\
= ~ & \TailP'_0 + \sum_{i = 2}^{N - 1} \epsilon \cdot (1 + \epsilon)^{i - 2}
\cdot \TailP'_{i-1} \\
= ~ & \TailP'_0 + \sum_{i = 1}^{N - 2} \epsilon \cdot (1 + \epsilon)^{i - 1}
\cdot \TailP'_{i}
~ \leq ~ \Approximation' ~ ,
\end{align*}
} 
Fix $\TailP'_i = \Probability(X_i \geq (1 + \epsilon)^{i})$ for every
$0 \leq i \leq N - 1$ and define
$$
\Approximation'
= \TailP'_0
+ \sum_{i = 1}^{N - 1} \epsilon \cdot (1 + \epsilon)^{i - 1} \cdot
\TailP'_i ~ .
$$
Clearly, $\TailP_i' \leq \TailP_i$ for every $0 \leq i \leq N - 1$, thus
$\Approximation' \leq \Approximation$.
On the other hand, for every $1 \leq i \leq N - 1$, we have
$$
\TailP'_{i - 1}
= \Probability(X_{i - 1} \geq (1 + \epsilon)^{i - 1})
\geq \Probability(X_{i} \geq (1 + \epsilon)^{i - 1})
\geq \Probability(X \geq (1 + \epsilon)^{i})
= \TailP_{i} ~ ,
$$
where the last inequality is due to
Proposition~\ref{proposition:ShrinkingShortEdges}.
Since, $\TailP'_0 = \TailP_0$, we get
\LongVersion
\DetailsEquationApproximationBoundTwo{}
therefore
\begin{equation} \label{equation:ApproximationBound2}
\Approximation /(1 + \epsilon)
\leq \Approximation'
\leq \Approximation ~ .
\end{equation}
\LongVersionEnd
\ShortVersion
\begin{equation} \label{equation:ApproximationBound2}
\Approximation /(1 + \epsilon)
\leq \Approximation'
\leq \Approximation
\end{equation}
(refer to Appendix~\ref{appendix:DetailsEquationApproximationBoundTwo} for
details).
\ShortVersionEnd
So, our next goal is to approximate $\Approximation'$.

\paragraph{Relying on local approximators.}
\Comment{
\subsection{Relying on local approximators}
} 
Consider some $n$-vertex BCW graph $H$ and denote the random variable that
takes on $\Property(H)$ by $X_H$.
Let $\delta, \hat{\delta} > 0$ be some performance parameters.
Recall that Karger's FPRAS for ATNR can be used to generate a real
$\Estimator^{> 0}$ that serves as a $(\delta \cdot \Probability(X_{H} > 0),
\hat{\delta})$-approximator of $\Probability(X_{H} > 0)$ in time
$\Polynomial(n, 1 / \delta, \log(1 / \hat{\delta}))$ (see
Section~\ref{section:Preliminaries}).
In Section~\ref{section:ProcedureEstimate} we present
Procedure~\ProcedureEstimate{} that given some $x > 0$, runs in time
$\Polynomial(n, 1 / \delta, \log(1 / \hat{\delta}))$ and returns a real
$\Estimator^{\geq x}$ that serves as a $(\delta \cdot \Probability(X_{H} > 0),
\hat{\delta})$-approximator of $\Probability(X_{H} \geq x)$.

Set the performance parameters $\delta \leftarrow \left( \frac{\epsilon}{n (1
+ \epsilon)} \right)^2$ and $\hat{\delta} \leftarrow 1 / (8 N)$.
For $i = 0, 1, \dots, N - 1$, we invoke Karger's FPRAS and our
Procedure~\ProcedureEstimate{} on $H \leftarrow G_i$ with $x \leftarrow (1 +
\epsilon)^i$ to produce the \emph{local approximators} $\Estimator^{> 0}
\rightarrow \Estimator_{i}^{> 0}$ and $\Estimator^{\geq x} \rightarrow
\Estimator_{i}^{\geq x}$.
We then define
$$
\TailP''_i =
\left\{
\begin{array}{ll}
\Estimator_{0}^{> 0} & \text{ if } i = 0; \\
\max\{ \Estimator_{i}^{\geq x}, \Estimator_{i + \kappa}^{> 0} \} & \text{ if
} 0 < i < N - \kappa; \\
\Estimator_{i}^{\geq x} & \text{ if } N - \kappa \leq i \leq N - 1
\end{array}
\right.
\quad \text{and} \quad
\Approximation''
= \TailP''_0
+ \sum_{i = 1}^{N - 1} \epsilon \cdot (1 + \epsilon)^{i - 1} \cdot
\TailP''_i ~ .
$$
How well does $\Approximation''$ approximates $\Approximation'$?
\LongVersion 
In attempt to answer this question, we first establish the following lemma.
\LongVersionEnd 
\ShortVersion 
In attempt to answer this question, we first establish the following lemma
whose proof is deferred to
Appendix~\ref{appendix:AdditionalProofsFpras}
\ShortVersionEnd 

\begin{lemma} \label{lemma:TailProbabilitiesAreClose}
With probability at least $3 / 4$, we have
\LongVersion \\ \LongVersionEnd
(a) $|\TailP'_i - \TailP''_i| \leq \delta \TailP'_0$ for all $0 \leq i \leq
\kappa$; and
\LongVersion \\ \LongVersionEnd
(b) $|\TailP'_i - \TailP''_i| \leq \delta \TailP'_{i - \kappa}$ for all
$\kappa < i \leq N - 1$.
\end{lemma}
\newcommand{\ProofLemmaTailProbabilitiesAreClose}{
By the choice of $\hat{\delta} = 1 / (8 N)$, we may use a union bound argument
and conclude that the inequalities
\begin{align}
& |\Probability(X_{H} > 0) - \Estimator^{> 0}| \leq \delta \cdot
\Probability(X_{H} > 0) \label{equation:EstimationGt} \\
& |\Probability(X_{H} \geq x) - \Estimator^{\geq x}| \leq \delta \cdot
\Probability(X_{H} > 0) \label{equation:EstimationGeq}
\end{align}
hold (simultaneously) for all $N$ invocations of Karger's FPRAS and
Procedure~\ProcedureEstimate{} with probability at least $3 / 4$;
the remainder of the proof is conditioned on that event.
By the definition of the shrunk graphs, we have
\begin{equation} \label{equation:LowIndices}
\Probability(X_i > 0)
= \Probability(X_0 \geq 1)
= \TailP'_0
\quad
\text{for every } 0 \leq i \leq \kappa
\end{equation}
and
\begin{equation} \label{equation:HighIndices}
\Probability(X_i > 0)
= \Probability(X_i \geq (1 + \epsilon)^{i - \kappa})
\leq \Probability(X_{i - \kappa} \geq (1 + \epsilon)^{i - \kappa})
= \TailP'_{i - \kappa}
\quad
\text{for every } \kappa < i \leq N - 1 ~ .
\end{equation}
Combining (\ref{equation:LowIndices}) and (\ref{equation:EstimationGt})
implies that $|\TailP'_0 - \TailP''_0| \leq \delta \TailP'_0$ as required for
$i = 0$.

Next, (\ref{equation:EstimationGeq}) guarantees that
\begin{equation} \label{equation:SpecificEstimationGeq}
|\Probability(X_i \geq (1 + \epsilon)^i) - \Estimator_{i}^{\geq x}|
\leq \delta \cdot \Probability(X_i > 0)
\end{equation}
for every $0 \leq i \leq N - 1$, while (\ref{equation:EstimationGt})
guarantees that
\begin{align}
\Estimator_{i + \kappa}^{> 0}
\leq & \Probability(X_{i + \kappa} > 0) + \delta \cdot \Probability(X_{i +
\kappa} > 0) \nonumber \\
\leq & \Probability(X_i \geq (1 + \epsilon)^i) + \delta \cdot
\Probability(X_{i + \kappa} > 0) \label{equation:ShrunkPhases} \\
\leq & \Probability(X_i \geq (1 + \epsilon)^i) + \delta \cdot \Probability(X_i
> 0) \label{equation:SpecificEstimationGt}
\end{align}
for every $0 \leq i < N - \kappa$, where (\ref{equation:ShrunkPhases}) is due
to (\ref{equation:HighIndices}) and (\ref{equation:SpecificEstimationGt})
follows from the definition of the shrunk graphs.
By combining (\ref{equation:SpecificEstimationGeq}) and
(\ref{equation:SpecificEstimationGt}), it follows that for every $0 < i < N -
\kappa$, we have
$$
\Probability(X_i \geq (1 + \epsilon)^i) - \delta \cdot \Probability(X_i > 0)
\leq \max\{ \Estimator_{i}^{\geq x}, \Estimator_{i + \kappa}^{> 0} \}
\leq \Probability(X_i \geq (1 + \epsilon)^i) + \delta \cdot \Probability(X_i
> 0) ~ ,
$$
or in other words,
$$
|\Probability(X_i \geq (1 + \epsilon)^i) - \max\{ \Estimator_{i}^{\geq x},
\Estimator_{i + \kappa}^{> 0} \}| \leq \delta \cdot \Probability(X_i > 0) ~ .
$$
This yields the desired $|\TailP'_i - \TailP''_i| \leq \delta \TailP'_0$ for
every $0 < i \leq \kappa$ due to (\ref{equation:LowIndices});
and $|\TailP'_i - \TailP''_i| \leq \delta \TailP'_{i - \kappa}$ for every
$\kappa < i < N - \kappa$ due to (\ref{equation:HighIndices}).

It remains to show that $|\TailP'_i - \TailP''_i| \leq \delta \TailP'_{i -
\kappa}$ for every $N - \kappa \leq i \leq N - 1$.
This is a direct consequence of (\ref{equation:SpecificEstimationGeq}) and
(\ref{equation:HighIndices}).
} 
\LongVersion 
\begin{proof}
\ProofLemmaTailProbabilitiesAreClose{}
\end{proof}
\LongVersionEnd 

We are now ready to complete the analysis.
Lemma~\ref{lemma:TailProbabilitiesAreClose} guarantees that
\LongVersion
\begin{align*}
|\Approximation' - \Approximation''|
= ~ & \left| \TailP'_0 - \TailP''_0 + \sum_{i = 1}^{N - 1} \epsilon \cdot (1 +
\epsilon)^{i - 1} \cdot (\TailP'_i - \TailP''_i) \right| \\
\leq ~ & |\TailP'_0 - \TailP''_0| + \sum_{i = 1}^{N - 1} \epsilon \cdot (1 +
\epsilon)^{i - 1} \cdot |\TailP'_i - \TailP''_i| \\
\leq ~ & \delta \TailP'_0 \left[ 1 + \sum_{i = 1}^{\kappa} \epsilon \cdot (1
+ \epsilon)^{i - 1} \right]
+ \sum_{i = \kappa + 1}^{N - 1} \epsilon \cdot (1 + \epsilon)^{i - 1} \cdot
\delta \TailP'_{i - \kappa} \\
= ~ & \delta \TailP'_0 \left[ 1 + \epsilon \cdot \frac{(1 + \epsilon)^{\kappa}
- 1}{\epsilon} \right]
+ \delta \cdot (1 + \epsilon)^{\kappa} \, \sum_{i = \kappa + 1}^{N - 1}
\epsilon \cdot (1 + \epsilon)^{i - \kappa - 1} \cdot \TailP'_{i - \kappa} \\
= ~ & \delta \cdot (1 + \epsilon)^{\kappa} \cdot \left[ \TailP'_0 + \sum_{i =
1}^{N - \kappa - 1} \epsilon \cdot (1 + \epsilon)^{i - 1} \cdot \TailP'_i
\right] \\
\leq ~ & \delta \cdot (1 + \epsilon)^{\kappa} \cdot \Approximation'
~ \leq ~ \delta \cdot \frac{n^2 (1 + \epsilon)^2}{\epsilon} \cdot
\Approximation'
\end{align*}
\LongVersionEnd
\ShortVersion
$|\Approximation' - \Approximation''|
\leq |\TailP'_0 - \TailP''_0| + \sum_{i = 1}^{N - 1} \epsilon \cdot (1 +
\epsilon)^{i - 1} \cdot |\TailP'_i - \TailP''_i|
\leq \delta \TailP'_0 \left[ 1 + \sum_{i = 1}^{\kappa} \epsilon \cdot (1
+ \epsilon)^{i - 1} \right] + \sum_{i = \kappa + 1}^{N - 1} \epsilon \cdot (1
+ \epsilon)^{i - 1} \cdot \delta \TailP'_{i - \kappa}$
with probability at least $3 / 4$, and thus
\begin{align*}
|\Approximation' - \Approximation''|
~ \leq ~ & \delta \TailP'_0 \left[ 1 + \epsilon \cdot \frac{(1 +
\epsilon)^{\kappa} - 1}{\epsilon} \right] + \delta \cdot (1 +
\epsilon)^{\kappa} \, \sum_{i = \kappa + 1}^{N - 1} \epsilon \cdot (1 +
\epsilon)^{i - \kappa - 1} \cdot \TailP'_{i - \kappa} \\
= ~ & \delta \cdot (1 + \epsilon)^{\kappa} \cdot \left[ \TailP'_0 + \sum_{i =
1}^{N - \kappa - 1} \epsilon \cdot (1 + \epsilon)^{i - 1} \cdot \TailP'_i
\right]
~ \leq ~ \delta \cdot (1 + \epsilon)^{\kappa} \cdot \Approximation'
~ \leq ~ \delta \cdot \frac{n^2 (1 + \epsilon)^2}{\epsilon} \cdot
\Approximation'
\end{align*}
\ShortVersionEnd
with probability at least $3 / 4$.
By the choice of $\delta = \left( \frac{\epsilon}{n (1 + \epsilon)}
\right)^2$, we conclude that $|\Approximation' - \Approximation''| \leq
\epsilon \Approximation'$.
Theorem~\ref{theorem:FprasBcw} follows by combining the last inequality with
(\ref{equation:ApproximationBound1}) and
(\ref{equation:ApproximationBound2}).

\section{Procedure~\ProcedureEstimate{}}
\label{section:ProcedureEstimate}
In this section we present and analyze Procedure~\ProcedureEstimate{}.
Let $G$ be some $n$-vertex BCW graph and denote the random variable that takes
on $\Property(G)$ by $X$.
Consider some positive real $x$ and performance parameters $\epsilon,
\hat{\epsilon} > 0$.
Given $G$ and $x$, Procedure~\ProcedureEstimate{} runs in time
$\Polynomial(n, 1 / \epsilon, \log(1 / \hat{\epsilon}))$ and outputs an
$(\epsilon \cdot \Probability(X > 0), \hat{\epsilon})$-approximator of
$\Probability(X \geq x)$.

Recall that requirement~\RequirementContract{} implies that edges of zero
weight can be contracted without affecting the value of $\Property$.
This means that we may contract every edge $e \in \Edges(G)$ such that
$W_{e}^{\Heavy} = W_{e}^{\Light} = 0$ without affecting $\Probability(X > 0)$
and $\Probability(X \geq x)$.
Indeed, in what follows we assume that $W_{e}^{\Heavy} > 0$ for all edges $e
\in \Edges(G)$.

Fix $E = \Edges(G)$.
It will be convenient to partition the edges in $E$ according to their light
phases to $E^0 = \{ e \in E \mid W_{e}^{\Light} = 0 \}$ and to $E - E^0 = \{ e
\in E \mid W_{e}^{\Light} > 0 \}$.
Let $G^0$ be the restriction of $G$ to the edges in $E^0$.
Recall that the event $X = \Property(G) > 0$ occurs if and only if
$\Property(G^0) > 0$ (see Section~\ref{section:Preliminaries});
denote the probability of this event by $\ProbZero$.

Fix $c = (5 + \sqrt{17}) / 2 \approx 4.56$. 
If $\ProbZero$ is sufficiently large, specifically, at least $n^{-c}$, then
the desired approximation can be obtained by a direct Monte Carlo method.
Indeed, Theorem~\ref{theorem:MonteCarloChernoff} guarantees that a
Monte Carlo method with $O (\log(1 / \hat{\epsilon}) n^{c} / \epsilon^2) =
\Polynomial(n, 1 / \epsilon, \log(1 / \hat{\epsilon}))$ trials suffices to
generate an $(\epsilon \cdot \ProbZero, \hat{\epsilon})$-approximator of
$\Probability(X \geq x)$.
(The random variable for which we apply the Monte Carlo method is simply the
indicator of the event $X \geq x$.)
This applies in particular to the case where $G^0$ is disconnected which means
that $\ProbZero = 1$.
Therefore in what follows we may assume that $\ProbZero < n^{-c}$ and in
particular, that $G^0$ is connected.
Note that if $\ProbZero$ is extremely small (e.g., exponentially small in
$n$), then the above Monte Carlo method requires too many samples in order to
obtain an $(\epsilon \cdot \ProbZero, \hat{\epsilon})$-approximator of
$\Probability(X \geq x)$.

\paragraph{Dealing with small $\ProbZero$.}
\Comment{
\subsection{Dealing with small $\ProbZero$}
}
How do we efficiently generate an $(\epsilon \cdot \ProbZero,
\hat{\epsilon})$-approximator of $\Probability(X \geq x)$ when $\ProbZero$ is
small?
For that purpose we introduce the real valued random variable $Y$ which is
defined over the probability space $E - E^0$ by mapping each instance
$\Instance : E - E^0 \rightarrow \{\Heavy, \Light\}$ to $\Probability(X \geq x
\mid \Instance)$, namely, $\Instance$ is mapped to the probability that
$\Property(G)$ is at least $x$ conditioned on $\Instance$.
This can be viewed as decomposing the probability space $E$ into the Cartesian
product of the probability spaces $E - E^0$, from which $\Instance$ is chosen,
and $E^0$, over which $\Probability(X \geq x \mid \Instance)$ is defined.
A crucial observation here is that
\begin{align*}
\Expectation[Y]
= & \sum_{\Instance : E - E^0 \rightarrow \{\Heavy, \Light\}}
\Probability(\Instance) \cdot \Probability(X \geq x \mid \Instance) \\
= & \sum_{\Instance : E - E^0 \rightarrow \{\Heavy, \Light\}} \Probability(X
\geq x \wedge \Instance) \\
= & ~ \Probability(X \geq x) ~ ,
\end{align*}
hence our goal is to provide a good approximation for $\Expectation[Y]$.
Another important observation is that
$$
\Probability(X \geq x \mid \Instance)
~ \leq ~ \Probability(X > 0 \mid \Instance)
~ = ~ \Probability(X > 0)
~ = ~ \ProbZero
$$
for every instance $\Instance : E - E^0 \rightarrow \{\Heavy, \Light\}$
(recall that the event $X > 0$ does not depend on the probability space $E -
E^0$), therefore $Y \in [0, \ProbZero]$ with probability $1$.

Fix $k = 2 \ln(4 / \hat{\epsilon}) / \epsilon^2$ and repeat the following
process for $j = 1, \dots, k$.
Choose some instance $\Instance_j : E - E^0 \rightarrow \{\Heavy, \Light\}$
with probability $\Probability(\Instance_j)$ (this can be easily generated by
randomly choosing the phase of each edge in $E - E^0$ independently of all
other edges) and let $Y_j = \Probability(X \geq x \mid \Instance_j)$;
in other words, $Y_j$ is a random sample of $Y$.
Unfortunately, we do not know how to efficiently compute the exact value of
$Y_j$ for a given instance $\Instance_j$.
Instead, we will generate an \emph{approximate sample} $Y'_j$ which is an
$(\epsilon \cdot \ProbZero / 2, \hat{\epsilon} / (2 k))$-approximator of $Y_j$.

We will soon explain how the $Y'_j$s are generated, but first let us explain
how they are employed to obtain the desired $(\epsilon \cdot \ProbZero,
\hat{\epsilon})$-approximator of $\Expectation[Y]$.
Let $\bar{Y} = \sum_{j = 1}^{k} Y_j / k$ and $\bar{Y}' = \sum_{j = 1}^{k} Y'_j
/ k$.
Theorem~\ref{theorem:MonteCarloHoeffding} guarantees that $\bar{Y}$ is an
$(\epsilon \cdot \ProbZero / 2, \hat{\epsilon} / 2)$-approximator of
$\Expectation[Y]$.
By Proposition~\ref{proposition:ApproximatorsAdditivity}, we conclude that
$\bar{Y}'$ is an $(\epsilon \cdot \ProbZero / 2, \hat{\epsilon} /
2)$-approximator of $\bar{Y}$.
Therefore Proposition~\ref{proposition:ApproximatorsTransitivity} implies that
$\bar{Y}'$ is an $(\epsilon \cdot \ProbZero, \hat{\epsilon})$-approximator of
$\Expectation[Y] = \Probability(X \geq x)$ as required.

\paragraph{Generating the approximate samples.}
\Comment{
\subsection{Generating the approximate samples}
}
It remains to present the process through which the approximate samples $Y'_j$
are generated (recall that each approximate sample should be an $(\epsilon
\cdot \ProbZero / 2, \hat{\epsilon} / (2 k))$-approximator of $\Probability(X
\geq x \mid \Instance)$ for some given instance $\Instance : E - E^0
\rightarrow \{\Heavy, \Light\}$).
The technique we use for this process is an extension of Karger's technique
\cite{Karg99}.
In order to simplify the description of this process, we first assume that
there exists some real $p$ such that $p_e = p$ for all edges $e \in E^0$
(all these edges have zero light phase, however their heavy phases may vary).
This assumption is removed later on.

Given some compact cut $C$ of $G^0$, let $\AllHeavy(C)$ denote the event that
all edges in $\Edges(C) \subseteq E^0$ take on their heavy (positive) phases.
Let $C$ be some min cut of $G^0$ and let $\chi = |\Edges(C)|$ be its size.
Since $\Probability(\AllHeavy(C)) = p^{\chi}$, the assumption that $\ProbZero
< n^{-c}$ implies that $p^{\chi} < n^{-c}$.
The following two theorems are established in \cite{Karg99} for the case of
$2$-way cuts.
Building upon the recent bound of Berend \& Tassa on the Bell number
\cite{BT10}, we extend them to (compact) $r$-way cuts for all $r \geq 2$
\LongVersion
simultaneously.
\LongVersionEnd
\ShortVersion
simultaneously (proofs are deferred to
Appendix~\ref{appendix:AdditionalProofsProcedureEstimate}).
\ShortVersionEnd

\begin{theorem} \label{theorem:FewSmallCuts}
For every real $\alpha \geq 1$, there are less than $13 n^{2 \alpha}$ compact
cuts of size at most $\alpha \chi$ in $G^0$.
Moreover, these cuts can be enumerated in expected time $\tilde{O} (n^{2
\alpha})$.
\end{theorem}
\newcommand{\ProofTheoremFewSmallCuts}{
The theorem is established by presenting a random process that generates each
compact cut of size at most $\alpha \chi$ in $G^0$ with probability greater
than $\frac{1}{13} n^{-2 \alpha}$.
Observe first that if an  $r$-way cut $C$ satisfies $|\Edges(C)| \leq \alpha
\chi$, then $r$ must be at most $2 \alpha$ as otherwise there exists some
cluster $U$ of $C$ with less than $\chi$ edges crossing between $U$ and
$\Vertices(G^0) - U$, in contradiction to the assumption that $\chi$ is the
size of a min cut of $G^0$.

Fix $k = \lceil 2 \alpha \rceil$.
Our random process first performs random edge contractions in $G^0$ until $k$
vertices $v_1, \dots, v_k$ remain in the graph (cf. Section~2.2.1 in
\cite{Karg99});
each vertex $v_i$ corresponds to some subset $V_i \subseteq \Vertices(G^0)$ so
that $\{V_1, \dots, V_k\}$ is a partition of $\Vertices(G^0)$ (the subgraph
induced on $G^0$ by $V_i$ is connected).
We then take $P = \{W_1, \dots, W_{\ell}\}$, $1 \leq \ell \leq k$, to be a
partition of $\{v_1, \dots, v_k\}$ chosen uniformly at random out of the $B_k$
possible partitions of $\{v_1, \dots, v_k\}$, where $B_k$ is the
$k^{\text{th}}$ \emph{Bell number}.
The cut $\widehat{C} = \{U_1, \dots, U_{\ell}\}$ generated by our random
process is defined by setting $U_j = \bigcup \{V_i \mid v_i \in W_j\}$ for $j
= 1, \dots, \ell$.
It is important to note that $\widehat{C}$ is not necessarily a compact cut,
however, if $C$ is any compact cut of size at most $\alpha \chi$ in $G^0$,
then it can be generated by our random process.
Our goal in the remainder of this proof is to show that $C$ is indeed
generated with probability greater than $\frac{1}{13} n^{-2 \alpha}$.

Karger \cite{Karg99} shows that the probability that none of the edges
crossing $C$ is contracted during the random edge contractions is at least
$$
\left( 1 - \frac{2 \alpha}{n} \right) \left( 1 - \frac{2 \alpha}{n - 1}
\right) \cdots \left( 1 - \frac{2 \alpha}{k + 1} \right)
= \frac{{k \choose 2 \alpha}}{{n \choose 2 \alpha}} ~ ,
$$
where generalized binomial coefficients\footnote{
Generalized binomial coefficients are a generalization of the standard
binomial coefficients ${x \choose y}$ to non-integral $x$ and $y$.
This generalization is based on replacing the factorial in the standard
definition with the Gamma function.
Many of the identities and bounds that hold for the standard binomial
coefficients also hold in the generalized case, including the bounds
$\left( \frac{x}{y} \right)^y \leq {x \choose y} \leq \left( \frac{e x}{y}
\right)^y$.
} are used when $2 \alpha$ is not an integer.
It remains to prove that $\frac{{k \choose 2 \alpha}}{{n \choose 2 \alpha}}
B_{k}^{-1} > \frac{1}{13} n^{-2 \alpha}$.
Indeed,
\begin{align}
\frac{{k \choose 2 \alpha}}{{n \choose 2
\alpha}} B_{k}^{-1}
\geq ~ & \left( \frac{k}{2 \alpha} \right)^{2 \alpha} \left( \frac{2 \alpha}{e
n} \right)^{2 \alpha} B_{k}^{-1} \nonumber \\
> ~ & \left( \frac{k}{e n} \right)^{2 \alpha} \left( \frac{\ln (k + 1)}{0.792
k} \right)^{k} \label{equation:BerendTassa} \\
> ~ & n^{-2 \alpha} \cdot \frac{1}{k} \left( \frac{\ln (k + 1)}{0.792 e}
\right)^{k} \nonumber ~ ,
\end{align}
where inequality~\ref{equation:BerendTassa} is due to Berend \& Tassa
\cite{BT10}.
The assertion follows as $\frac{1}{k} \left( \frac{\ln (k + 1)}{0.792 e}
\right)^{k} > \frac{1}{13}$ when $k \geq 2$.
} 
\LongVersion
\begin{proof}
\ProofTheoremFewSmallCuts{}
\end{proof}
\LongVersionEnd

\begin{theorem} \label{theorem:LowProbabilityForAllLargeCuts}
For every real $\alpha \geq 1$, the probability that there exists some compact
cut $C$ of size at least $\alpha \chi$ in $G^0$ such that $\AllHeavy(C)$
occurs is $O (n^{-\alpha \eta})$, where $\eta$ is defined by fixing $p^{\chi}
= n^{-(2 + \eta)}$.
\end{theorem}
\newcommand{\ProofTheoremLowProbabilityForAllLargeCuts}{
Let $C_1, \dots, C_t$ be the compact cuts of size at least $\alpha \chi$ and
for each $1 \leq i \leq t$, let $\chi_i = |\Edges(C_i)|$.
We assume without loss of generality that $\alpha \chi \leq \chi_1 \leq \cdots
\leq \chi_t$.
Denote $p_i = p^{\chi_i} = \Probability(\AllHeavy(C_i))$ and consider some
real $\beta \geq 1$.
By Theorem~\ref{theorem:FewSmallCuts}, there are less than $13 n^{2 \beta}$
compact cuts of size at most $\beta \chi$.
It follows that $\chi_{13 n^{2 \beta}}$ must be greater than $\beta \chi$.

We shall bound the probability that $\AllHeavy(C_i)$ occurs for some (at least
one) $1 \leq i \leq t$ by bounding the sum $\sum_{i = 1}^{t} p_i$.
The first $t' = 13 n^{2 \alpha + 1 / \ln(n)}$ terms are bounded simply by
observing that
$$
\sum_{i = 1}^{t'} p_i
\leq 13 n^{2 \alpha + 1 / \ln(n)} \cdot p^{\alpha \chi}
= 13 e \cdot n^{2 \alpha} \cdot n^{-\alpha (2 + \eta)}
= 13 e \cdot n^{-\alpha \eta} ~ .
$$
Thus it remains to bound the remaining $t - t'$ terms.

Given some $\beta \geq \alpha$, we write $s = 13 n^{2 \beta + 1 / \ln(n)}$ and
conclude that $\chi_s > \left( \beta + \frac{1}{2 \ln(n)} \right) \chi =
\frac{\ln(s) - \ln(13)}{2 \ln n} \cdot \chi$.
Therefore
$$
p_s
< \left( p^{\chi} \right)^{\frac{\ln(s) - \ln(13)}{2 \ln n}}
= \left( n^{-(2 + \eta)} \right)^{\frac{\ln(s) - \ln(13)}{2 \ln n}}
= \left( e^{\frac{\ln(s) - \ln(13)}{2}} \right)^{-(2 + \eta)}
= s^{-(1 + \eta / 2)} \cdot 13^{1 + \eta / 2} ~ .
$$
Summing over all $i > t'$, we get
\begin{align*}
\sum_{i > t'} p_i
~ < ~ & 13^{1 + \eta / 2} \cdot \sum_{s > t'} s^{-(1 + \eta / 2)} \\
\leq ~ & 13^{1 + \eta / 2} \cdot \int_{t'}^{\infty} s^{-(1 + \eta / 2)} d s
\\
= ~ & 13^{1 + \eta / 2} \cdot \left( -(\eta / 2) \cdot
s^{-\eta / 2} \bigg\vert_{13 n^{2 \alpha + 1 / \ln(n)}}^{\infty} \right) \\
= ~ & 13^{1 + \eta / 2} \cdot (\eta / 2) \cdot 13^{-(\eta / 2)} \cdot
n^{-\alpha \eta} \cdot e^{-\eta / 2} \\
\leq ~ & 13 \cdot n^{-\alpha \eta} ~ .
\end{align*}
The assertion follows.
} 
\LongVersion
\begin{proof}
\ProofTheoremLowProbabilityForAllLargeCuts{}
\end{proof}
\LongVersionEnd

Notice that the compact cuts addressed in Theorems \ref{theorem:FewSmallCuts}
and \ref{theorem:LowProbabilityForAllLargeCuts} may have an arbitrary number
of clusters, but their size is compared to $\alpha \chi$, where $\chi$ is the
size of the smallest $2$-way cut in $G^0$.
This point is crucial for the validity of the arguments.

Write $p^{\chi} = n^{-(2 + \eta)}$.
We must have $\eta > c - 2$ as $p^{\chi} < n^{-c}$.
Fix $\alpha = \frac{c - 1 + \ln(1 / \epsilon) / \ln(n)}{2}$ and let
$\mathcal{C}$ be the collection of all compact cuts of size at most $\alpha
\chi$ in $G^0$.
By Theorem~\ref{theorem:FewSmallCuts}, $\mathcal{C}$ consists of $O (n^{2
\alpha}) = O (n^{c - 1} / \epsilon)$ compact cuts that can be enumerated in
expected time $\tilde{O} (n^{c - 1} / \epsilon)$.

Given some sub-collection $\mathcal{B} \subseteq \mathcal{C}$, let
$$
\psi(\mathcal{B})
= \Probability \left( \bigvee_{C \in \mathcal{B}} \AllHeavy(C)
\right)
$$
be the probability that all crossing edges of some (at least one) cut in
$\mathcal{B}$ take on their heavy phases.
Theorem~\ref{theorem:LowProbabilityForAllLargeCuts} guarantees that
$0 \leq \ProbZero - \psi(\mathcal{C}) \leq \gamma n^{-\alpha \eta}$ for some
universal constant $\gamma$.
The choice of $c = (5 + \sqrt{17}) / 2$ and of $\alpha = \frac{c - 1 + \ln(1 /
\epsilon) / \ln(n)}{2}$ and the assumption that $\eta > c - 2$ ensure that
$\gamma n^{-\alpha \eta} \leq \epsilon n^{-(2 + \eta)} / 4 = \epsilon p^{\chi}
/ 4$ as long as $(4 \gamma)^4 \leq (1 / \epsilon)^{c - 4}$, which yields the
following corollary.

\begin{corollary} \label{corollary:LargeCutsDontMatter}
The probability that there exists some compact cut $C \notin \mathcal{C}$ such
that $\AllHeavy(C)$ occurs is at most $\epsilon p^{\chi} / 4 \leq \epsilon
\cdot \ProbZero / 4$.
\end{corollary}

Consider some instance $\Instance : E - E^0 \rightarrow \{\Heavy, \Light\}$.
Our goal is to efficiently generate an $(\epsilon \cdot \ProbZero / 2,
\hat{\epsilon} / (2 k))$-approximator of $\Probability(X \geq x \mid
\Instance)$.
For a given compact $r$-way cut $C = \{U_1, \dots, U_r\}$ of $G^0$, we
construct the graph $G_{C, \Instance}$ as follows.
The vertex set of $G_{C, \Instance}$ is $\Vertices(G_{C, \Instance}) = \{u_1,
\dots, u_r\}$.
For every edge $e \in E^0$ with one endpoint in the cluster $U_i$ and the
other in the cluster $U_j$, $i \neq j$, we add an edge $(u_i, u_j)$ to
$\Edges(G_{C, \Instance})$ whose weight is $W_{e}^{\Heavy}$.
In addition, for every edge $e \in E - E^0$ with one endpoint in the cluster
$U_i$ and the other in the cluster $U_j$, $i \neq j$, we add an edge $(u_i,
u_j)$ to $\Edges(G_{C, \Instance})$ whose weight is $W_{e}^{\Instance(e)}$.
The following observation is due to requirement~\RequirementContract{}.

\begin{observation} \label{observation:SameDiameter}
Conditioned on the instance $\Instance : E - E^0 \rightarrow \{\Heavy,
\Light\}$, and on the event that the set of edges in $E_0$ that take on their
heavy phases induces the compact $r$-way cut $C$ on $G^0$, we have
$\Property(G_{C, \Instance}) = X$.
\end{observation}

Let $\mathcal{B}_{\Instance}$ be the collection of all compact cuts $C$ of
$G^0$ such that $\Property(G_{C, \Instance}) \geq x$.
Observation~\ref{observation:SameDiameter} implies that $\Probability(X \geq
x \mid \Instance) = \psi(\mathcal{B}_{\Instance})$.
By Corollary~\ref{corollary:LargeCutsDontMatter}, we know that
$\psi(\mathcal{B}_{\Instance} - \mathcal{C}) \leq \epsilon \cdot \ProbZero /
4$, and hence $\psi(\mathcal{B}_{\Instance}) - \epsilon \cdot \ProbZero / 4
\leq \psi(\mathcal{B}_{\Instance} \cap \mathcal{C}) \leq
\psi(\mathcal{B}_{\Instance})$.
Consequently, it suffices to generate an $(\epsilon \cdot \ProbZero / 4,
\hat{\epsilon} / (2 k))$-approximator of $\psi(\mathcal{B}_{\Instance} \cap
\mathcal{C})$.

\paragraph{Probabilistic DNF satisfiability.}
\Comment{
\subsection{Probabilistic DNF satisfiability}
}
The approximation of $\psi(\mathcal{B}_{\Instance} \cap \mathcal{C})$ is
performed by the method of Karp, Luby, and Madras \cite{KL85,KLM89} for
approximating the probability that a formula in disjunctive normal form (DNF)
is satisfied.
Given some DNF formula $\phi$, and given the probability $q_i$ that $x_i$ is
assigned to true for each variable $x_i$ (independently of all other
variables), the method of Karp et al. generates a $(\delta \cdot q(\phi),
\hat{\delta})$-approximator of the probability $q(\phi)$ that $\phi$ is
satisfied in time $O (|\phi| \log(1 / \hat{\delta}) / \delta^2)$, where
$|\phi|$ stands for the size of the formula (number of literals).

Cast into that framework, the event $\bigvee_{C \in \mathcal{B}_{\Instance}
\cap \mathcal{C}} \AllHeavy(C)$ is encoded as a DNF formula whose
variables correspond to whether or not the edges in $E^0$ take on their heavy
phases and whose clauses correspond to the cuts in $\mathcal{B}_{\Instance}
\cap \mathcal{C}$.
Such a DNF formula has $|\mathcal{B}_{\Instance} \cap \mathcal{C}| \leq
|\mathcal{C}| = O (n^{c - 1} / \epsilon)$ clauses, each with at most $n$
literals.
Therefore an $(\epsilon \cdot \psi(\mathcal{B}_{\Instance} \cap \mathcal{C}) /
4, \hat{\epsilon} / (2 k))$-approximator of $\psi(\mathcal{B}_{\Instance} \cap
\mathcal{C})$, which also serves as an $(\epsilon \cdot \ProbZero / 4,
\hat{\epsilon} / (2 k))$-approximator of $\psi(\mathcal{B}_{\Instance} \cap
\mathcal{C})$ since $\psi(\mathcal{B}_{\Instance} \cap \mathcal{C}) \leq
\ProbZero$, can be generated in time
$O (\log(k / \hat{\epsilon}) n^{c} / \epsilon^{3})
= \Polynomial(n, 1 / \epsilon, \log(1 / \hat{\epsilon}))$.

\paragraph{Varying heavy phase probabilities.}
\Comment{
\subsection{Varying heavy phase probabilities}
}
Recall that in attempt to simplify the description of the process that
generates approximate samples of the random variable $Y$, we assumed that $p_e
= p$ for all edges $e \in E^0$.
\LongVersion 
We now turn to lift this assumption.
The technique we use here is essentially identical to that used by Karger
\cite{Karg99} for a similar purpose; we describe it for completeness.
\LongVersionEnd 
\ShortVersion 
The technique we use to lift this assumption is essentially identical to that
used by Karger \cite{Karg99} for a similar purpose and we defer its
description to Appendix~\ref{appendix:ProofVaryingHeavyPhaseProbabilities}.
\ShortVersionEnd 

\newcommand{\ProofVaryingHeavyPhaseProbabilities}{
The BCW graph $G^0$ with varying heavy phase probabilities $0 < p_e < 1$ is
transformed into a BCW graph $H$, $\Vertices(H) = \Vertices(G^0)$, with
uniform heavy phase probabilities $p = 1 - \theta$ for some sufficiently small
$\theta > 0$.
For each edge $e = (u, v) \in E^0$ with heavy phase probability $0 < p_e < 1$,
we introduce a \emph{bundle} of $k_e = \ln(1 / p_e) / \theta$ parallel $(u,
v)$ edges in $H$ with the same heavy and light phases as $e$ (recall that the
light phase of $e$ is zero).
The probability that all $k_e$ edges in this bundle take on their heavy phase
is $(1 - \theta)^{\ln(1 / p_e) / \theta}$ which converges to $p_e$ as $\theta
\rightarrow 0$.
By requirement~\RequirementParallel{}, it is sufficient to generate approximate
samples for the random variable $Y$ with respect to the graph $H$ in the limit
as $\theta \rightarrow 0$.
The technique we introduced earlier in this section is suitable for that as
$H$ has uniform heavy phase probabilities.
In particular it is sufficient to enumerate all the small compact cuts $C$ of
$H$, identify those inducing $\Property(G_{C, \Instance}) \geq x$ for a
given instance $\Instance : E - E^0 \rightarrow \{\Heavy, \Light\}$, and then
approximate the probability that all crossing edges of at least one of them
take on their heavy phases.

Note that changing the parameter $\theta$ scales the size of cuts in
$H$ without changing their relative sizes.
We construct a graph $H'$, $\Vertices(H') = \Vertices(G^0)$, with positive
costs on the edges by assigning cost $\ln(1 / p_e)$ to each edge $e \in E^0$
(the phases of the edges in $E^0$ are ignored in the context of $H'$).
The small cost cuts in $H'$ correspond to the small sized cuts in $H$;
they can be enumerated by known techniques.

Given the small cuts in $H$ that induce $\Property(G_{C, \Instance}) \geq x$,
we have to approximate the probability, as $\theta \rightarrow 0$, that all
crossing edges of at least one of them take on their heavy phases.
We already argued that this is exactly the probability that all crossing edges
of at least one of the corresponding cuts in $G^{0}$ take on their heavy
phases.
Approximating this probability is done as before by constructing the
appropriate DNF formula and employing the method of Karp et
al. \cite{KL85,KLM89}.
} 
\LongVersion 
\ProofVaryingHeavyPhaseProbabilities{}
\LongVersionEnd 

\LongVersion 
\section{Transforming RW graphs into BCW graphs}
\label{section:TransformingRwToBcw}
\LongVersionEnd 
\newcommand{\SectionTransformingRwToBcw}{
So far, we have developed an approximation for
$\Expectation[\Property(G)^{k}]$, where $G$ is a BCW graph.
In this section we are interested in extending our algorithm to the (general)
case of RW graphs.
This extension relies on an efficient transformation of any RW graph $G$ into
a BCW graph $G'$ such that the random variable $\Property(G')$ is
stochastically equivalent to the random variable $\Property(G)$ and $|G'| = O
(|G|)$.
This is similar to a method presented by Mirchandani \cite{Mirc76} (see also
\cite{BCP95}).

Let $G$ be an arbitrary RW graph and consider some edge $e \in \Edges(G)$.
Recall that there exists some positive integer $m(e)$ and some non-negative
phases $W_{e}^{1}, \dots, W_{e}^{m(e)}$ and probabilities $p_{e}^{1}, \dots,
p_{e}^{m(e)}$, where $\sum_{i = 1}^{m(e)} p_{e}^{i} = 1$, such that
$\Weight(e) = W_{e}^{i}$ with probability $p_{e}^{i}$ (independently of all
other edges).
We assume without loss of generality that if $m(e) > 2$, then the phases of
$e$ are distinct (identical phases can be merged into one) and ordered so that
$W_{e}^{1} < \cdots < W_{e}^{m(e)}$.

The BCW graph $G'$ is obtained by replacing every edge $e = (u, v) \in
\Edges(G)$ such that $m(e) > 2$ with $m(e) - 1$ parallel edges $e_1, \dots,
e_{m(e) - 1} = (u, v) \in \Edges(G')$, each having exactly two phases.
The heavy phases of all new edges are set to be $W_{e}^{m(e)}$.
The light phase of $e_i$ is set to be $W_{e_i}^{\Light} = W_{e}^{i}$ for
every $1 \leq i \leq m(e) - 1$.
The probabilities $p(e_1), \dots, p(e_{m(e) - 1})$ are designed to guarantee
that the random variable $M_e = \min\{ \Weight(e_1), \dots, \Weight(e_{m(e) -
1}) \}$ in $G'$ is stochastically equivalent to the random variable
$\Weight(e)$  in $G$.
This is achieved by setting
$$
p(e_i) = 1 - \frac{p_{e}^{i}}{1 - \sum_{j = 1}^{i - 1} p_{e}^{j}}
= \frac{1 - \sum_{j = 1}^{i} p_{e}^{j}}{1 - \sum_{j = 1}^{i - 1} p_{e}^{j}} ~ .
$$
Indeed, for every $1 \leq i \leq m(e)$, we have
$$
\Probability(M_e = W_{e}^{i})
= \prod_{j = 1}^{i - 1} p(e_j) \cdot (1 - p(e_i))
= \prod_{j = 1}^{i - 1} \frac{1 - \sum_{l = 1}^{j} p_{e}^{l}}{1 - \sum_{l =
1}^{j - 1} p_{e}^{l}} \cdot \frac{p_{e}^{i}}{1 - \sum_{j = 1}^{i - 1}
p_{e}^{j}}
= p_{e}^{i} ~ ,
$$
where the last equation holds by telescoping.
Requirement~\RequirementParallel{} implies that $\Property(G')$ is
stochastically equivalent to $\Property(G)$.
Theorem~\ref{gtheorem:Fpras} now follows (at least when $k = 1$) from
Theorem~\ref{theorem:FprasBcw}.
} 
\LongVersion 
\SectionTransformingRwToBcw{}
\LongVersionEnd 

\LongVersion 
\section{Hardness}
\label{section:Hardness}
\LongVersionEnd 
\newcommand{\SectionHardness}{
In this section we prove that the problem of computing the expected diameter
of an RW graph is \SharpP{}-hard.
The problem remains \SharpP{}-hard even when restricted to identically
distributed weighted graphs.
Our line of arguments immediately implies that computing the radius of an
identically distributed weighted graph with respect to a designated vertex is
also \SharpP{}-hard.

Hardness is established by reduction from the most basic variant of the
\emph{two terminal network reliability (TTNR)} problem defined as follows.
On input connected graph $G$ and two vertices $s, t \in \Vertices(G)$, the
goal is to compute the probability $P$ that $s$ and $t$ become disconnected
when each edge in $\Edges(G)$ is removed with probability $1 / 2$
independently of all other edges.
The \SharpP{}-hardness of TTNR is established by Valiant \cite{Vali79}.
Since the support of $P$ consists of integer multiples of $2^{-m}$, where
$m = |\Edges(G)|$, we conclude that it is \SharpP{}-hard to approximate $P$ to
within a one-sided additive error of $\epsilon$ for any $\epsilon < 2^{-m}$.

Given a graph $G$ with two vertices $s, t \in \Vertices(G)$ as input of TTNR,
we construct an identically distributed weighted graph $G'$ with parameters $p
= 1 / 2$, $W^{\Heavy} = 1$, and $W^{\Light} = 0$.
$G'$ is obtained from $G$ by adding a new edge $e = (s, t)$ and augmenting the
resulting graph with two simple paths, one connecting $s$ to the new vertex
$s'$ and the other connecting $t$ to the new vertex $t'$.
Each new simple path consists of $k = \Theta (m)$ new vertices.
The reduction is cast in the following lemma.

\begin{lemma}
Let $D$ be the random variable that takes on $\Diameter(G')$.
Then $P \leq 2 (\Expectation[D] - k) < P + 2^{-m}$.
\end{lemma}
\begin{proof}
\ShortVersion 
\begin{AvoidOverfullParagraph}
\ShortVersionEnd 
Let $D'$ be the random variable that takes on $\Distance_{G'}(s', t')$.
By definition, we know that $D' \leq D$ with probability $1$.
We shall take $k$ to be sufficiently large so that Chernoff's inequality
implies that $\Probability(\Distance_{G'}(s', s) < n \vee \Distance_{G'}(t,
t') < n) < 2^{-(m + 1)} / n$, where $n = |\Vertices(G)|$.
By the construction of $G'$, it follows that $\Probability(D > D') < 2^{-(m +
1)} / n$.
Since $D - D' < n$, we conclude that $0 \leq \Expectation[D] -
\Expectation[D'] < 2^{-(m + 1)}$.
\ShortVersion 
\end{AvoidOverfullParagraph}
\ShortVersionEnd 

By the linearity of expectation, we have
$$
\Expectation[D']
= \Expectation[\Distance_{G'}(s', s)] + \Expectation[\Distance(s, t)] +
\Expectation[\Distance_{G'}(t, t')]
= k + \Expectation[\Distance(s, t)] ~ ,
$$
where the last term can be rewritten as
\begin{align*}
\Expectation[\Distance(s, t)]
= & ~ \Expectation[\Distance(s, t) \mid \Weight(e) = 1] \cdot
\Probability(\Weight(e) = 1) + \Expectation[\Distance(s, t) \mid \Weight(e) =
0] \cdot \Probability(\Weight(e) = 0) \\
= & ~ \Expectation[\Distance(s, t) \mid \Weight(e) = 1] / 2 ~ .
\end{align*}
The assertion is established by arguing that $\Expectation[\Distance(s, t)
\mid \Weight(e) = 1] = P$.
Indeed, when $\Weight(e) = 1$, then $\Distance_{G'}(s, t) \in \{0, 1\}$.
The argument holds since the $\Distance_{G'}(s, t) = 0$ instances
(respectively, the $\Distance_{G'}(s, t) = 1$ instances) of the probability
space $\Edges(G')$ correspond to the instances of TTNR in which $s$ and $t$
remain connected (resp., become disconnected).
\end{proof}
} 
\LongVersion 
\SectionHardness{}
\LongVersionEnd 

\section{Conclusions}
\label{section:Conclusions}
We study the setting of graphs whose edge weights are independent random
variables and show that for the wide family of distance-cumulative weighted
graph properties, the problem of computing the $k^{\text{th}}$ moment admits
an FPRAS.
Computing the expectation (i.e., the first moment) for example is difficult
when the variance is large, and hence too many samples are required for the
Monte Carlo method in order to take into account low probability
events\footnote{
Events occurring with probability which is low with respect to the limited
computational resources of our algorithm are still expected to occur in
practice if the network in hand (say, the Internet) is being used very
frequently. 
} that may drastically affect the expectation.
Our technique does not guarantee a (multiplicative) approximation for the
\emph{$k^{\text{th}}$ central moment} (and in particular, the variance) when
this is close to zero, however, it does provide us with the ability to decide
if the $k^{\text{th}}$ central moment is close to zero.

There are still some fundamental weighted graph properties which are not
distance-cumulative, and hence cannot be dealt with via our technique, such as
the shortest $(s, t)$-path and the weight of a maximum matching.
It is also natural to consider the directed analogue of randomly weighted
graphs and in particular, various network flow problems.
Another aspect that calls for further research, once encoding issues are
resolved, is that of continuous distributions for the edge weights.

\section*{Acknowledgment}
We would like to thank Noga Alon for helpful discussions.

\ShortVersion 
\clearpage
\renewcommand{\thepage}{}
\ShortVersionEnd 


\clearpage

\pagenumbering{roman}
\appendix

\renewcommand{\theequation}{A-\arabic{equation}}
\setcounter{equation}{0}

\begin{center}
\textbf{\large{APPENDIX}}
\end{center}

\ShortVersion 
\section{Additional proofs for Section~\ref{section:Fpras}}
\label{appendix:AdditionalProofsFpras}

\begin{proof}[Proof of Proposition~\ref{proposition:ShrinkingShortEdges}]
\ProofPropositionShrinkingShortEdges{}
\end{proof}

\begin{proof}[Proof of Lemma~\ref{lemma:TailProbabilitiesAreClose}]
\ProofLemmaTailProbabilitiesAreClose{}
\end{proof}

\subsection{Explaining inequality~(\ref{equation:ApproximationBound2})}
\label{appendix:DetailsEquationApproximationBoundTwo}

The inequality stands as
\DetailsEquationApproximationBoundTwo{}
and since $\Approximation' \leq \Approximation$.

\section{Additional proofs for Section~\ref{section:ProcedureEstimate}}
\label{appendix:AdditionalProofsProcedureEstimate}

\begin{proof}[Proof of Theorem~\ref{theorem:FewSmallCuts}]
\ProofTheoremFewSmallCuts{}
\end{proof}

\begin{proof}[Proof of Theorem~\ref{theorem:LowProbabilityForAllLargeCuts}]
\ProofTheoremLowProbabilityForAllLargeCuts{}
\end{proof}

\section{Varying heavy phase probabilities}
\label{appendix:ProofVaryingHeavyPhaseProbabilities}

\ProofVaryingHeavyPhaseProbabilities{}

\section{Transforming RW graphs into BCW graphs}
\label{section:TransformingRwToBcw}

\SectionTransformingRwToBcw{}

\section{Hardness}
\label{section:Hardness}

\SectionHardness{}

\ShortVersionEnd 

\section{High critical ratio}
\label{appendix:CriticalRatio}

Consider the identically distributed weighted graph $G$ consisting of $2$
vertices and $m$ parallel edges connecting them, where each edge is of weight
$1$ with probability $1 / 2$; and of weight $2^{2 m}$ otherwise.
Let $X$ denote the random variable that takes on the diameter of $G$.
It is easy to verify that $\Expectation[X] \approx 2^m$, while $\Variance[X]
\approx 2^{3 m}$, so the critical ratio here is roughly $2^m$.
Indeed, a Monte Carlo method with significantly less than $2^m$ samples would
probably estimate the expected diameter of $G$ to be $1$ which is an awful
approximation.

\end{document}